\documentclass[prb,twocolumn,showpacs]{revtex4}
\usepackage{graphicx}
\usepackage{mathbbol}
\usepackage{amsmath}
\usepackage{amssymb}
\usepackage{epstopdf}
\usepackage{slashed}

\begin{document}
\title{Interlayer correlated fractional quantum Hall state in the $\nu=4/5$ bilayer system}
\author{Hao Wang}
\affiliation{Shenzhen Institute for Quantum Science and Engineering and Department of Physics, Southern University of Science and Technology, Shenzhen 518055, China}

\author{Alexander Seidel}
\affiliation{Department of Physics, Washington University, St.
Louis, Missouri 63130, USA}

\author{Kun Yang}
\affiliation{Physics Department and National High Magnetic Field Laboratory, Florida State University, Tallahassee, Florida 32310, USA}

\author{Fu-Chun Zhang}
\affiliation{Kavli Institute for Theoretical Sciences, and CAS Center for Topological Quantum Computation, University of Chinese Academy of Sciences, Beijing 100190, China}

\begin{abstract}
We perform exact diagonalization studies for fractional quantum Hall states at filling factor $4/5$ in a bilayer system, on a torus with various aspect ratios and angles. We find that in the absence of tunneling, two weakly coupled $2/5$-layers undergo a phase transition into an interlayer-correlated regime, which is also Abelian with the five-fold degeneracy on the torus. In the limit of zero layer separation, this phase becomes a singlet in the pseudospin variable describing the layer degree-of-freedom. By studying the Chern-number matrix, we show that the $K$-matrix describing the interlayer-correlated regime requires matrix dimension larger than two and  this regime is in particular not described by a Halperin state. A detailed analysis of possible $4\times 4$ $K$-matrices having the requisite symmetries and quantum numbers shows that there is only one equivalence class of such matrices. A model wave function representing this universality class is constructed. The role of separate particle number conservation in both layers is discussed, and it is argued that this additional symmetry allows for the further distinction of two different symmetry protected Abelian phases in the interlayer correlated regime. Interlayer tunneling breaks this symmetry, and can drive the system into a single-layer regime when strong enough. A qualitative phase diagram in the tunneling-layer separation parameter space is proposed based on our numerical results.
\end{abstract}

\pacs{73.43.Cd, 73.43.Nq, 73.43.-f}
\maketitle

\section{Introduction}

Bilayer quantum Hall systems host extremely rich physics, due to the electron's (additional) internal degree of freedom associated with the layer where it resides, as well as the competition between inter- and intra-layer electron-electron interactions. In addition to double quantum well systems, a single wide quantum well can also support a bilayer regime, depending on system parameters. Recent experiments observed a fractional quantum Hall (FQH) state at Landau filling $\nu=4/5$ in such wide quantum well systems.\cite{YangLiu} Obvious candidates include the single-layer $\nu=4/5$ state which is the particle-hole conjugate of the Laughlin $\nu=1/5$ state in the single layer regime, and two weakly coupled $\nu=2/5$ states (one for each layer) in the bilayer regime, with weak interlayer interaction and correlation. Much more interesting is the bilayer regime in which interlayer interaction strength is comparable to intralayer interaction, and interlayer electron-electron correlation cannot be neglected. In this regime several theoretical model states have been suggested, some of which may be non-Abelian.\cite{scarola:2001,BarkeshliWen,Balram:2015} Motivated by these experimental and theoretical developments, we perform detailed numerical studies of a clean bilayer quantum Hall system at total Landau filling $\nu=4/5$, with equal population of the two layers and in most cases, without interlayer tunneling. The effects of interlayer tunneling will be briefly addressed in Sec. \ref{sec:tunneling}.

\section{Model and Numerical Results}
\label{model}

In our numerical calculations, we consider a bilayer electron system subject to a magnetic field $B$ perpendicular to the two-dimensional plane. We use the torus geometry with two-dimensional basis vectors $\mathbf{L_x}$, $\mathbf{L_y}$ spanning the unit cell and having an aspect angle $\theta$ between them. Unless otherwise stated, numerical results presented in this paper correspond to $L_x=L_y$ and $\theta=\pi/2$, i.e., square unit cell, although we will also consider several other cell geometries. There is an integer number of magnetic flux quanta $N_{\phi}=L_x L_y \sin{\theta}/2 \pi \ell^2$ going through the cell, where the magnetic length $\ell=\sqrt{\hbar c/eB}$ is chosen as the length unit and the energy is in units of $e^2/4\pi \varepsilon \ell$. To reduce the size of the Hilbert space, we carry out our calculation within every pseudomomentum sector $\mathbf{K}=(K_x,K_y)$\cite{rezayi:2000}. The magnetic field is assumed to be strong enough so that electrons can be regarded as spin-polarized or spinless, \cite{morf:1998,rezayi:2000} and confined to the lowest Landau level. The Coulomb interactions are then projected onto the lowest Landau level\cite{rezayi:2000}, and can be written in the form
\begin{eqnarray}
  H_{c}=
    \frac{1}{N_{\phi}} \sum_{\mathbf{q\ne0,\alpha,\beta}} V_{\alpha\beta}(\mathbf{q}) e^{-q^2/2}
\sum_{i<j}e^{i \mathbf{q} \cdot (\mathbf{R}_{\alpha,i}-\mathbf{R}_{\beta,j})}.
\label{hamiltonian}
\end{eqnarray}
Here, $\alpha(\beta)=1,2$ are indices referring to the two layers. The momentum $\mathbf{q}=(q_x,q_y)$ takes discrete values suitable for the given unit cell and $\mathbf{R}_{\alpha,i}$ is the guiding center coordinate of the $i$-th electron on the layer $\alpha$. $V_{\alpha,\alpha}(\mathbf{q})=1/q$ and $V_{\alpha \ne \beta}(\mathbf{q})=e^{-qd}/q$ are the Fourier transforms of the intralayer and interlayer Coulomb interactions, respectively, and $d$ is the layer separation. In the present work, we consider the layers with zero width and performed exact diagonalization to obtain the energy spectrum and eigenstates for a total number of electrons $N_e$ up to 12.

\begin{figure}[t]
\centerline{\includegraphics [width=3.4 in] {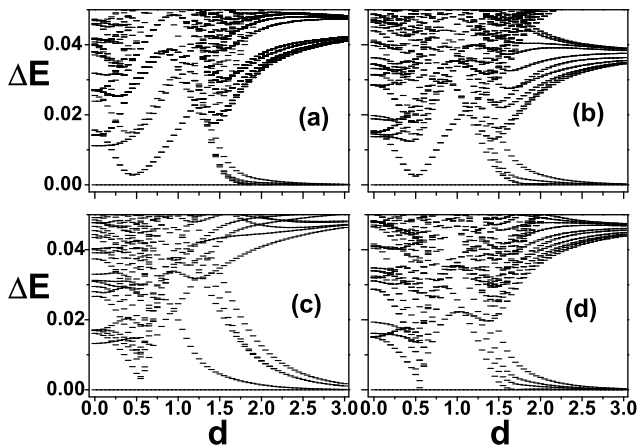}}
\caption{Low-lying excitation spectra versus layer separation for $\nu=4/5$ bilayer system with $N_e=12$ and cell geometries of (a) aspect angle $\theta=60^{\circ}$, aspect ratio $r=L_x/L_y=1.0$ (hexagonal cell); (b) $\theta=80^{\circ}$, $r=1.0$; (c) $\theta=90^{\circ}$, $r=0.8$; (d) $\theta=90^{\circ}$, $r=1.0$ (square cell).} \label{Egap}
\end{figure}

Figure \ref{Egap} plots excitation energy spectra as functions of layer separation, for various unit cell geometries at a system size of $N_e=12$ with both layers equally occupied  and without interlayer tunneling. We start our discussion in the large-$d$ region, where five lowest energy levels at the pseudomomentum sector (0,0) form a nearly degenerate group that is clearly separated from higher energy states by a gap. Together with the trivial 5-fold center of mass (COM) degeneracy on the torus, this indicates the existence of a 25-fold (nearly degenerate) ground state (GS) manifold. In the $d \rightarrow \infty$ limit where two layers completely decouple, we note that these manifold states become exactly degenerate.  Each $\nu=2/5$ layer can be understood as a well-studied 2/5 FQH state, carrying its own 5-fold COM degeneracy.

\begin{figure}[t]
\centerline{\includegraphics [width=3 in] {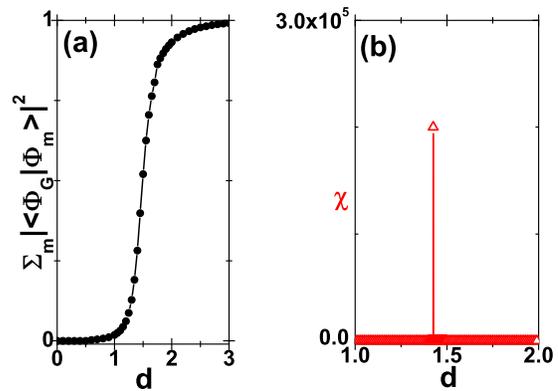}}
\caption{(Color online) $N_e=12$ system at a given GS pseudomomentum sector (0,0): (a) Sum of squared wave function overlap between ground state and model states vs. layer separation. (b) Susceptibility of group fidelity for the five lowest states vs. layer separation.} \label{wfoverlap}
\end{figure}

Our numerical calculation shows that the system in the large-$d$ region is actually connected to the decoupled-layer state in the $d \rightarrow \infty$ limit, namely a state of two weakly-coupled 2/5 FQH layers. A separate calculation has confirmed that, as $d$ grows to infinity, the excitation gap above the GS manifold, $\Delta E_s=E_6-E_5$, approaches the gap of a $\nu=2/5$ single layer, and the GS energy of the system approaches twice that of the single layer. The connection can also be exhibited through a comparison of wave functions as shown in Fig. \ref{wfoverlap}(a), where a model state constructed from the single-layer 2/5 FQH states and the ground state $\Phi_G$ at pseudomomentum sector (0,0) of the $N_e=12$ system are considered. The single-layer 2/5 FQH system of $N_e=6$ has five degenerate ground states at sectors (0,3$m$) for $m=0,...,4$. Then the five basis states $\Phi_m$ for the bilayer system at sector (0,0) can be built as the direct product of these, i.e., $\Phi_m=(0,3m)\bigotimes(0,k'_m)$ with $k'_m=\textbf{Mod}(N_{\phi}-3m,N_{\phi})$. The model state of the bilayer system is considered as a linear combination in this basis. As shown in the plot, the sum of the squared wave function overlaps between $\Phi_G$ and these $\Phi_m$ does continuously evolve to unity in the $d \rightarrow \infty$ limit.

The 25-fold degeneracy is of topological origin\cite{WenAdvPhys:1995}. As long as the system stays in the weakly-coupled-layer phase, it should remain exact in the thermodynamic limit. However, this degeneracy could be lifted by finite-size effect for any $d < \infty$. As shown in Fig. \ref{Egap}, such lifting increases with decreasing $d$. Nearly for all systems at $d=d_{c2}\sim 1.5$, the lifting becomes comparable to the gap and mixing between states below and above the gap starts. Besides of this spectrum feature, the wave function overlap plot in Fig. \ref{wfoverlap}(a) also exhibits a sharp transition around $d_{c2}$. These observation suggest a quantum phase transition occurring at $d_{c2}$. To probe the existence of the phase transition at $d_{c2}$, we performed a fidelity test for the five lowest states $\psi_i$ ($i=1,..,5$) at pseudomomentum sector (0,0). The group fidelity $f$ and its corresponding susceptibility $\chi$ are defined as:
\begin{eqnarray}
  f&=&\frac{1}{5}\sum_{i,j=1}^5|\langle\psi_i(d-\delta)|\psi_j(d)\rangle|^2,\\
  \chi&=&(1-f)/\delta^2.
\label{fidelity}
\end{eqnarray}
The results show a sharp peak around $d_{c2}$, signifying the quantum phase transition occurring there.

To investigate the nature of the transition at $d_{c2}$, we turn to the spectrum with $d$ further decreasing. We note that the lowest energy level does not engage in any mixing with other levels, remaining separated by a gap except for $d=d_{c1}\sim 0.5$, where another transition occurs and we will discuss it later. Correspondingly, the GS degeneracy changes to become 5-fold below $d_{c2}$. Thus, the transition at $d_{c_2}$ can be interpreted as a transition between the large-$d$ phase of weakly coupled $2/5$ layers with 25-fold degenerate ground state, and a topologically different phase with 5-fold degenerate ground state at smaller $d$. Such a transition would also require excited states to become gapless in the thermodynamic limit. A softening of some exited states around $d\sim 1.5$ is clearly seen in the spectra. However, due to finite size effects, the mixing between these softened modes and (some of) the ground states still takes place at finite energy. At large $d$, the 25-fold ground state degeneracy signifies a stable topological phase, albeit one with relatively large degeneracy. It is not unexpected that moderate interlayer interactions (at intermediate $d$) suffice to lower this degeneracy, leading into a different topological phase, with, in this case, minimum COM-degeneracy of five. In the following discussion, we will refer to the regions with $d<d_{c_2}$ collectively as the interlayer-correlated FQH regime.

\begin{figure}[t]
\centerline{\includegraphics [width=3.4 in] {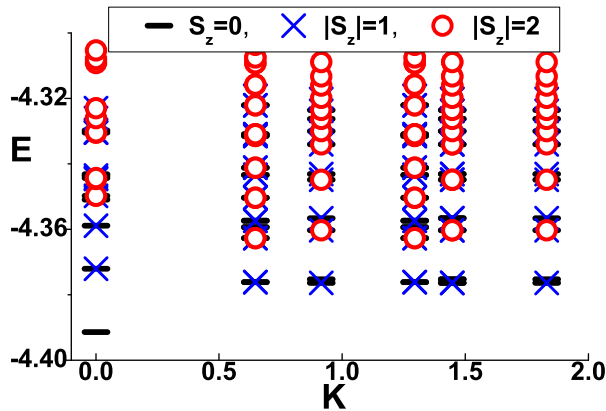}}
\caption{(Color online) Comparison of low-lying energy spectra at different  total pseudospins $|S_z|=0, 1, 2$ for $N_e=12$ system with $d=0$. } \label{pseudospin}
\end{figure}

As a first step to investigate the interlayer-correlated, small- to intermediate-$d$  region, we carry out numerical studies on the pseudospin excitation in the $d=0$ limit. In this case the inter- and intra-layer interactions are identical, and the system has a pseudospin SU(2)-symmetry when we identify the layer degree-of-freedom as a spin-$1/2$ pseudospin-index in the $S_z$-basis.  Denoting the number of electrons at each layer as $N_{\uparrow}=N_e/2+\Delta N$ and $N_{\downarrow}=N_e/2-\Delta N$, we then have a total of $S_z = \Delta N$ for the entire system. As a result of the SU(2)-symmetry, eigenstates can be labeled by SU(2) quantum numbers, where states with pseudospin $S$ come in multiplets with $-S \le S_z \le S$. As the numerical results in the Fig. \ref{pseudospin} show, the GS energy of a balanced-layer system ($\Delta N=0$ or $S_z=0$) is found to be the lowest while the GS energy of an imbalanced-layer system increases with $\Delta N$.  Furthermore, the GS energy of a $|S_z|=1$ system exactly matches with the energy of the first excited state of a balanced-layer system. These overlaps in low energy levels clearly indicate which pseudo-spin sector the excitation of the system belongs to. On the other hand, the GS energy of a $|S_z|=2$ system is higher. These observations suggest that the $\nu=4/5$ system at $d=0$ is a pseudospin-singlet state ($S=0$), while the lowest-energy excitations form a pseudospin triplet ($S=1$).

\section{Topological order of the interlayer-correlated states\label{TO}}

We now turn to a more in-depth study of the nature of the topological phase(s) in the interlayer-correlated regime. We first consider a generalized periodic boundary condition with twisted boundary phase angles $0\leq \theta_{\eta}^{\alpha} < 2\pi$ along $\eta=x,y$ directions in the layer $\alpha$. After a unitary transformation $\Psi=\exp[-i\sum_{\alpha}\sum_{i}((\theta_x^{\alpha}/L_x)x_i^{\alpha}+(\theta_y^{\alpha}/L_y)y_i^{\alpha})]\Phi$ , where the summation runs over all electrons of both layers, the resulting many-body wave function $\Psi$ once again satisfies (magnetic) periodic boundary conditions.  Many-body topological Chern numbers are then well-defined, and can be given as\cite{Thouless:1982,ShengPRL:2003}
\begin{equation}\label{cnm}
D\, C_{\alpha,\beta}=\frac{1}{2\pi}\mbox{Im} \sum_{i=1}^D\int\int d\theta_x^{\alpha}d\theta_y^{\beta}\left<\frac{\partial\Psi_i}{\partial\theta_y^{\beta}} \bigg|\frac{\partial\Psi_i}{\partial\theta_x^{\alpha}}\right>,
\end{equation}
where the integral is over a phase unit cell of $0\leq\theta_x^{\alpha},\theta_y^{\beta} \leq 2\pi$, $D$ is the ground state degeneracy, and $i$ is a label running over a basis of ground states. For the bilayer system with $\alpha(\beta)=1,2$, we thus have a $2 \times 2$ Chern number matrix (CNM).\cite{ShengPRL:2003} The off-diagonal matrix elements $C_{1,2}=C_{2,1}$ are relevant to the boundary phase averaged drag Hall conductance. Applying common (opposite) boundary phase on two layers, one can also get the boundary phase averaged charge (pseudospin) Hall conductance in units of $e^2/\hbar$ as $C_q=C_{1,1}+C_{2,2}+C_{1,2}+C_{2,1}$ ($C_s=C_{1,1}+C_{2,2}-C_{1,2}-C_{2,1}$). The CNM has been proposed equal to the inverse K-matrix for several Halperin hierarchy states in bilayer system and topological flat bands.\cite{chernmatrix:2017}

We calculated the CNM of the ground states for all the cell sizes, geometries, and ranges of the $d$ parameter considered. For $d<d_{c_1}$, the numerical results consistently give a $2\times2$ matrix
\begin{equation}
\label{metric}
C_1 = \left (
\begin{array}{cc}
-4/5 & 6/5 \\
6/5 & -4/5
\end{array}
\right ),
\end{equation}
whereas for $d_{c_1}<d\lesssim d_{c_2}$, the numerical results consistently give
\begin{equation}
\label{metric2}
C_2 = \left (
\begin{array}{cc}
6/5 & -4/5 \\
-4/5 & 6/5
\end{array}
\right ).
\end{equation}
These two regions are separated by a clear gap-closing feature in the spectrum at $d_{c_1} \approx 0.5$, where, even at finite system size, the gap closes exactly for some boundary condition, making the change of CNM possible. Since this transition does not involve a change in ground state degeneracy, we suspect that it is a symmetry protected topological phase transition related to the separate conservation of particle number in each layer, a point we will further elaborate on later. For $d\gtrsim d_{c_2}$, the lowest five energy states undergo mixing with 20 other states, and the assumption $D=5$ in Eq. \eqref{cnm} does not lead to consistent results. However, working with the 25 lowest energy states (setting $D=25$), at sufficiently large $d$ one recovers a CNM $C_{3}$ with diagonal elements equal to $2/5$ and zero off-diagonal elements, as expected for two separated 2/5-filling system. Results for $N_e=8$ particles are very similar, except with smaller intermediate region and earlier onset of the large-$d$ region, which we attribute to finite size effects.

We note that for the large-$d$ system with two separate layers at 2/5-filling, the inverse of the matrix $C_3$ does not correspond to a proper $2\times2$ K-matrix. This system is actually described by the $4\times4$ K-matrix
\begin{equation}
   K=\left(
   \begin{matrix}
      3 & 2 & 0 & 0 \\
       2 & 3 & 0& 0 \\
        0 & 0 & 3 & 2 \\
         0 & 0 & 2 & 3
   \end{matrix}
   \right)\,.
\end{equation}
Similarly, the inverses of both matrices $C_1$, $C_2$ do not lead to proper K-matrices, either. This suggests that the interlayer-correlated  regime is {\em not} described by a Halperin state. Nevertheless, our numerical findings, in particular the five-fold (minimal) torus-degeneracy, imply that the interlayer-correlated regime is Abelian. On general grounds, it should then be amenable to a description in terms of symmetric integer K-matrix, and a corresponding integer charge vector $q$. (For a review, see  Ref. \onlinecite{WenAdvPhys:1995}.) The fact that matrices $C_1$ and $C_2$  are not obtained as the inverses of proper K-matrices suggests that the dimension of the underlying K-matrix must be larger than 2, as in the case of $C_3$  in the large-$d$ limit.

A systematic search for a physically admissible K-matrix can be carried out as follows. Since the small-$d$ phase seems to be adiabatically connected to the singlet state  at $d=0$, and since the U(1)-invariance generated by (pseudo-)$S_z$ and (pseudo-)spin flip invariance remain symmetries for any $d$, the ground state must have $S_z=0$ and be invariant under spin flip for all $d$ below the phase transition at $d_{c_1}$. For the  K-matrix to exhibit this spin flip (or Ising) symmetry, we require it to be of even dimension $2n$ and commute with $\left ( \begin{smallmatrix}0 & \mathbb{1} \\ \mathbb{1} & 0\end{smallmatrix}\right)$, where $\mathbb{1}\equiv \mathbb{1}_{n\times n}$ is the $n\times n$-identity matrix. Furthermore, $|\det K|$ must equal the ground state  degeneracy of $5$, and  $q^t K^{-1}q$ must equal the filling factor of $4/5$. Finally, the components of $q$ must be co-prime for there to be trivial charge-$1$, electron-like excitations. There is then always a basis (the ``symmetric basis'') where all the components of $q$ are 1, and in this basis all diagonal components of $K$ must be odd, in order for the electron-like charge-$1$ excitations to be fermionic. Indeed, one finds no $n=1$ (two-dimensional) $K$-matrix that satisfies all these requirements, consistent with our earlier finding that the interlayer-correlated regime is not described by a Halperin state.

For $n=2$ the most general K-matrix thus described has the structure
\begin{equation}
   K=\left(
   \begin{matrix}
      a & b & d & e \\
       b & c & e& f \\
        d & e & a & b \\
         e & f & b & c
   \end{matrix}
   \right)\,,
\end{equation}
with $a$ and $c$ odd, and we take $q=(1,1,1,1)^t$. We carried out a brute force analysis of all such matrices with entries $-13\leq a,\dotsc,f\leq 13$. There are 447 such matrices satisfying $|\det K|=5$ and $q^t K^{-1}q=4/5$. However, all of these matrices have been found to be congruent via unimodular matrices that leave $q$ invariant. That is, different $K$, $K'$ in this set are all related via $K'=W K W^t$, where $W$ is an integer matrix with $|\det W|=1$ and $W q=q$. $K$ and $K'$ are then physically equivalent.\cite{WenAdvPhys:1995} These findings  suggest that, up to physical equivalence, there exists only one 4$\times$4 K-matrix with the requisite properties, with a suitable representative given by
\begin{equation}\label{Kfinal}
   K=\left(
   \begin{matrix}
      1 & 2 & 1 & 1 \\
       2 & 1 & 1 & 1 \\
        1 & 1 & 1 & 2 \\
         1 & 1 & 2 & 1
   \end{matrix}
   \right)\,.
\end{equation}
Any hypothetic K-matrix solving our problem while being {\em inequivalent} to Eq. \eqref{Kfinal} would need to have entries larger than 13 in the symmetric basis. Thus, Eq. \eqref{Kfinal} can certainly  be viewed as the simplest solution, suggesting that it is, at the very least, the most robust.

Note that Eq. \eqref{Kfinal} satisfies $\det K=+5$, with two positive and two negative eigenvalues, signifying two co-propagating and two counter-propagating bosonic edge modes. These are direct consequences of the bulk topological order encoded in the $K$ matrix, which can be revealed by inspecting edge excitation spectrum when studying disk geometry, or the entanglement spectrum of the ground state itself.

It is prudent to ask what the relation is between the $K$-matrix of Eq. \eqref{Kfinal} here and the Chern-number matrices discussed early. As mentioned, one of the defining features of Eq. \eqref{Kfinal} is the relation $q^t K^{-1}q=4/5$, which defines the filling factor, and, at the same time, the Hall conductance. As such, it is the sum of all matrix elements of the CNM. One may identify $q_1=(1,1,0,0)^t$ as the part of the charge vector associated to the upper layer, and $q_2=(0,0,1,1)^t$ the part of $q$ associated to the lower layer, such that $q=q_1+q_2$. We thus expect the CNM to be given by $C_{i,j}=q_i^t K^{-1}q_j$. Note that for Halperin states, $q_1=(1,0)^t$, $q_2=(0,1)^t$, which indeed reduces to CNM$=K^{-1}$. With this, the $K$-matrix of Eq. \eqref{Kfinal} gives matrix $C_2$ of Eq. \eqref{metric2}, corresponding to the region $d_{c_1}<d<d_{c_2}$. But it is inconsistent with the matrix $C_1$ of Eq. \eqref{metric} that we found for $d<d_{c_1}$. As already mentioned at the level of the $K$-matrix description, all $K$-matrices satisfying the constraints described here are equivalent in the topological sense\cite{WenAdvPhys:1995} (congruence via unimodular matrices), but not all of these matrices yield the same CNM via the above prescription. This may not be surprising, as the CNM is only well-defined in the presence of an additional U(1)-symmetry, and it is well appreciated that symmetries may {\em enrich}\cite{Wen2002} topological orders, i.e., they may lead to distinctions between phases that are otherwise topologically identical. Indeed, the equivalence class of $K$-matrices defined here does have representatives leading to the CNM $C_1$ of Eq. \eqref{metric}, the simplest being
\begin{equation}\label{Kfinal2}
   K'=\left(
   \begin{matrix}
      1 & 1 & 1 & 2 \\
       1 & 1 & 2 & 1 \\
        1 & 2 & 1 & 1 \\
         2 & 1 & 1 & 1
   \end{matrix}
   \right)\,.
\end{equation}
Indeed, Eqs. \eqref{Kfinal} and \eqref{Kfinal2} are congruent via the unimodular matrix
 \begin{equation}\label{Kultimate}
   W=\left(
   \begin{matrix}
      1 & 0 & 0 & 0 \\
       0 & 0 & 1 & 0 \\
        0 & 0 & 0 & 1 \\
         0 & 1 & 0 & 0
   \end{matrix}
   \right)\,.
\end{equation}
In particular, they describe the same edge physics, as well as bulk topological properties in terms of
quasi-particle content and statistics.

Finally we intend to provide in particular the $K$-matrix of Eq. \eqref{Kfinal} with a simple wave function interpretation: The diagonal blocks represent, in each layer, a particle-hole conjugate of the Laughlin-$1/3$ state, $\psi_{2/3}$. The layers are then coupled through a Jastrow-factor $\prod_{1\leq i,j\leq N} (z_i-w_j)$ (off-diagonal block), where  the $z_j$ ($w_j$) are the complex coordinates of the upper (lower) layer, thus
\begin{equation}\label{WF}
\begin{split}
 \psi(z_1,\dotsc,z_N,w_1, \dots, w_N)&= \\
 \prod_{1\leq i,j\leq N} (z_i-w_j)  \;\;\psi_{2/3}(z_1,\dotsc,&z_N) \psi_{2/3}(w_1,\dotsc, w_N)\,.
 \end{split}
\end{equation}
We note that a similar wave function form was also proposed by composite fermion theory.\cite{scarola:2001}

It is worth recalling that the $K$-matrix and charge vector $q$, by themselves, do not specify the topological shift\cite{WenZee:1992} $\cal S$ of the state, which is another important piece of the topological data and is defined in $N_\Phi= \nu^{-1} N_e -{\cal S}$ with $N_\Phi$ the number of flux quanta in spherical geometry. However, the variational wave function interpretation of Eq. \eqref{WF} does imply a definite topological shift as follows: One may convince oneself that the factor $\prod_{1\leq i,j\leq N} (z_i-w_j)$, applied to any (pseudo-)$S_z=0$ state, does not affect the topological shift. The topological shift of Eq. \eqref{WF} must therefore be the same as that of the $2/3$-state, which is ${\cal S}=0$. The shift ${\cal S}$ only requires rotational invariance in order to be well-defined, but not the separate conservation laws that lead to the distinction of the phases $d<d_{c_1}$ and $d_{c_1}<d<d_{c_2}$. Hence, one may expect the entire interlayer-correlated regime $d<d_{c_2}$ to have the same shift. Thus, to the extent that the wave function of Eq. \eqref{WF} is the correct description of the intermediate phase $d_{c_1}<d<d_{c_2}$, we expect this shift to be zero in all of the interlayer-correlated regime. A direct verification of this prediction on the sphere, as well as predictions on the edge and entanglement spectra, will provide further support to the topological order we have discussed here. We will leave these considerations for future work.

\section{Effects of interlayer tunneling}
\label{sec:tunneling}

We emphasize that well-defined CNMs of $C_1$ and $C_2$, separating two distinct regions, can be obtained only in the presence of separately conserved charges in the upper and lower layer, respectively. Under these circumstances, any change in CNM necessarily results in distinct regions separated by a critical point, as happens in the present case, where the CNM changes from $C_2$ to $C_1$ at $d_{c_1}$. On the other hand, if these separate conservation laws are violated via non-zero interlayer tunneling, one would expect that the region $d_{c_1}<d<d_{c_2}$ may be adiabatically connected to  the region $d<d_{c_1}$ without gap-closing. Shedding light on questions such as this, and to further explore the physics of a bilayer system at filling factor $4/5$, motivates us to explore an extended phase diagram including interlayer tunneling terms. Thus, we study the following model Hamiltonian with interlayer tunneling parameter $t$,
\begin{eqnarray}
  H'=H_{c}-t \cdot S_{x} + d \cdot S_{z}^2/N_\phi,
\label{ham-t}
\end{eqnarray}
where the total pseudospin $S_{x}=\sum_{i,\alpha \neq \beta}a_{i,\alpha}^\dag a_{i,\beta}/2$ with the creation (annihilation) operator $a_{i,\alpha}^\dag$ ($a_{i,\alpha}$) for the $i$-th lowest Landau level orbital in the $\alpha$-layer, and the last term is induced by an imbalance of electron number in the two layers with total pseudospin $S_{z}$.\cite{MacDonald:1990} This model will  allow us to investigate the phase evolution of the system under the influence of both the layer separation and interlayer tunneling, thus offering a more comprehensive understanding. Through our numerical calculation, we note that system GS is mostly located at the pseudomomentum sector (0,0) except for a small parameter region enclosed by the white dashed line in the diagram of Fig. 4. We interpret this to be a spin-density-wave regime,  which we argue removable via a perturbation on the Coulomb interaction, e.g., due to layer thickness. Therefore, in the following discussion we will focus on the ground state within the (0,0) sector.

\begin{figure}[t]
\centerline{\includegraphics [width=3.4 in] {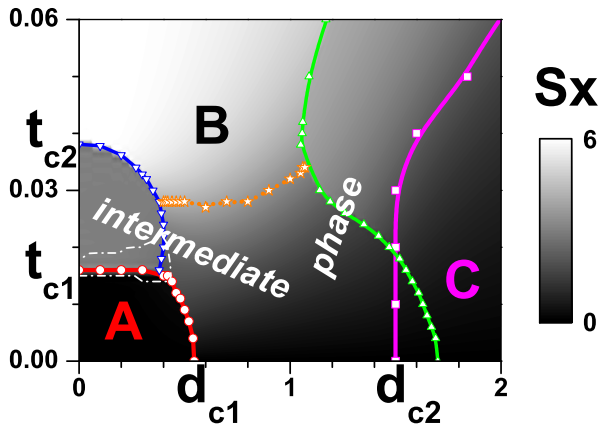}}
\caption{(Color online) Magnetization plot for the expectation value of total pseudospin $S_x$ at pseudomomentum sector (0,0) of $N_e=12$ system. The phase regions of A,B,C, intermediate phase, and their boundaries are described in the main text.} \label{phd}
\end{figure}

\begin{figure}[t]
\centerline{\includegraphics [width=3.4 in] {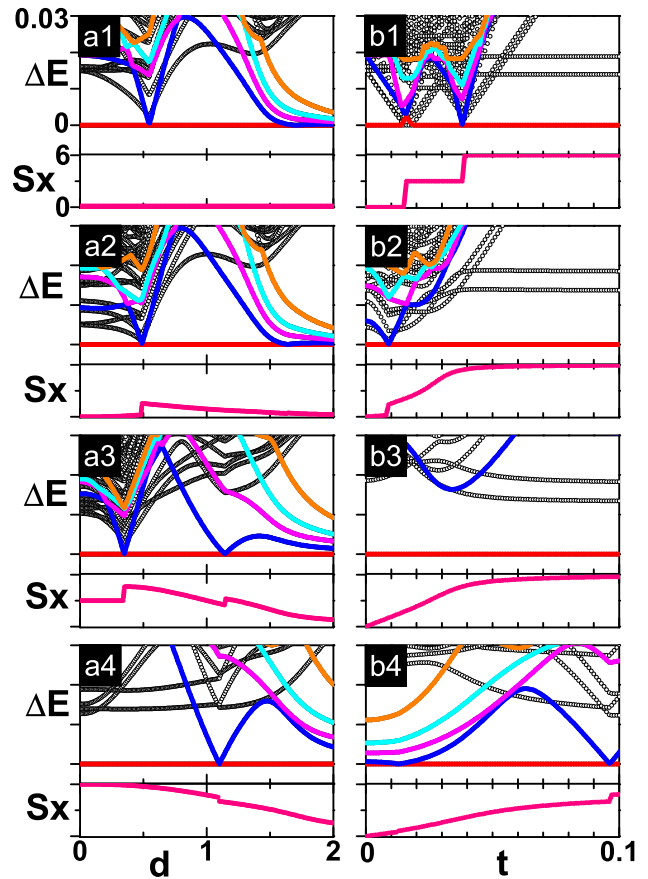}}
\caption{(Color online) Low lying excitation spectrum and expectation value of $S_x$ at pseudomomentum sector (0,0) for $N_e=12$ system. Five lowest energy levels at sector (0,0) are represented by colored curves in the spectra. Plots of (a1)-(a4) stand for system at selected interlayer tunneling parameters $t=0, 0.01, 0.03, 0.05$; (b1)-(b4) for system at selected layer separations $d=0, 0.5, 0.8, 1.6$.} \label{phd-bd}
\end{figure}

With the expectation value of total pseudospin $S_x$ at sector (0,0) depicted as a background grayscale, we have tentatively plotted an extended ``$t-d$'' phase diagram in Fig. 4 for the $N_e=12$ system with a square unit cell. The phase diagram of the $N_e=8$ system with a hexagonal cell has qualitatively similar features. The red/blue/green boundary lines in the diagram have been determined as locations of  a level-crossing involving the GS in the (0,0) sector. As is clearly seen from the background grayscale, these crossings are concomitant with jumps in the $S_x$-expectation value. {\em Therefore, these lines must be interpreted as lines of first order phase transitions.} Details regarding the discontinuity of $S_x$ in the (0,0) sector are shown in Fig. 5.

We first turn to the system with small interlayer tunneling $t \leq t_{c1}\approx 0.015$. As shown in Fig. 5(a2), spectral features are very similar to those previously discussed for $t=0$,  with two phase transitions as $d$ increases. Furthermore, we note that the system within the small $t$/small $d$ corner, which we termed the ``A'' phase, is adiabatically connected to the pseudospin singlet state at $t=0, d=0$, with its $S_x \approx 0$. This phase {\em continues} to be separated via a gap closing from an intermediate regime above the $d_{c_1}<d<d_{c_2}$ line. This is at first surprising, since, as we said initially, absent a well-defined $S_z$-quantum number (charge conservation in each layer), one would not expect these two regimes to be fundamentally distinct. However, since we have strong evidence that the red phase boundary is first order, there is no contradiction to the notion that these regions can be adiabatically connected in a larger Hamiltonian-space than presently considered. Moreover, the critical point at $d_{c_1}$, at which the CNM changes along the $d$-axis, serves as the natural second order terminal point for this line of first-order phase transitions.

We now turn to large layer separation $d$. Physically, even with moderate finite interlayer tunneling, the bilayer system in the limit $d\rightarrow\infty$ will enter the phase represented by two decoupled $2/5$ layers, which we denoted as ``C'' phase in the diagram. The (purple) boundary between this layer-decoupled zone and the layer-correlated zone in Fig. 4 is determined by scanning for the onset of a well-defined gap between the lowest five energy levels and the remainder in the spectrum. It can also be determined by the forming signal of the 25-fold degeneracy in wave function overlap and/or by the group fidelity as we discussed in the Sec. II. Based on numerical data, there exists another (green) boundary in the diagram, determined by the GS level-crossing and/or $S_x$-discontinuity. However, we note that the relative position between green and purple boundaries varies for $N_e=8$ system; the $S_x$ discontinuity at the bottom part of the green boundary is small as $t$ decreases and vanishes at $t=0$ as shown in Fig. 5 (a1-a2,b4). Also, as shown in Fig. 5(a1-a4), in the large $d$ region, the two crossing levels join the group of the five degenerate ground states (within the (0,0) sector), which are completely degenerate in the thermodynamic limit. These observations lead us to argue that the separate green and purple boundaries are due to finite-size effects and that these boundary lines would merge into a single boundary, at least below a certain $t$-value, in the thermodynamic limit.

\begin{figure}[t]
\centerline{\includegraphics [width=3.4 in] {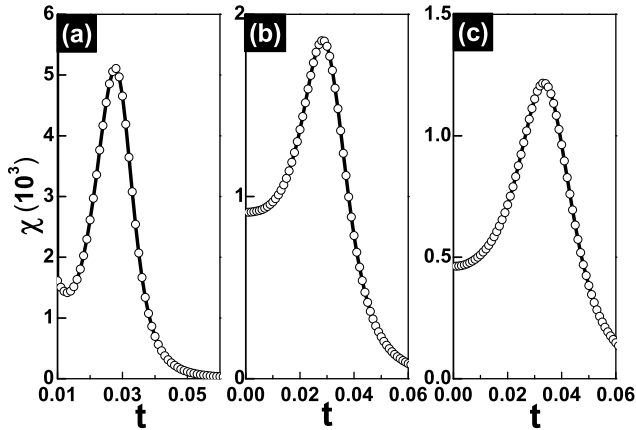}}
\caption{Ground state susceptibility at pseudomomentum sector (0,0) of $N_e=12$ system with selected layer separations: $d=$ (a) 0.5, (b) 0.8, (c) 1.05.} \label{phd-bd2}
\end{figure}

Finally, we look at the limit of large $t$. For finite $d$ and in the limit $t\rightarrow\infty$  the bilayer system effectively becomes a single-layer system as single particle orbitals must align with the effective field in the $x$-direction. We termed this phase the ``B'' phase. The transition from a two-layer system to a one-layer system can be exemplified in the $d=0$ case, where the symmetry is larger; in particular the $S_x$ symmetry remains intact. At $t\leq t_{c1}$, the system stays as a pseudospin singlet with $S_x=0$. Further increasing $t$, the system jumps to a finite $S_x$ and undergoes a sequence of GS level-crossing within the (0,0) sector, ending with an $S_x$-saturated phase at $t=t_{c2}\approx 0.038$, in which $S_x=N_e/2$ and the system enters the single layer regime described by
\begin{equation}
K = \left (
\begin{array}{cc}
1 & 0\\
0 & -5
\end{array}
\right ),
\end{equation}
corresponding to the particle-hole conjugate of Laughlin state at $\nu=1/5$. Although the discontinuity in $S_x$ and/or level-crossing in the spectrum are notable in both small-$d$ and large-$d$ regions, as shown by the blue and green lines in the Fig. 4, respectively, we note there is no such signal when we increase $t$ in the intermediate-$d$ region as shown in Fig. 5 (b2) and (b3). However, even along cuts in $t$ with $d$ in this region, we must expect a purely topological phase transition, without change in ground state degeneracy, as we attribute a different K-matrix to the $d_{c1}<d<d_{c2}$ interval on the $d$-axis (Sec. \ref{TO}). We have tried to determine the associated phase boundary in this intermediate region (the orange curve in the Fig. 4) using GS fidelity. As shown in Fig. 6, when we scan the GS fidelity/susceptibility at a given $d$, there is a minimum/maximum signal, which we take to be the transition point. This orange boundary merges with the upper parts of the blue boundary and green boundary at its two ends,  separating the one-layer phase with $S_x\approx N_e/2$ from the bilayer phase(s).
\begin{figure}[t]
\centerline{\includegraphics [width=3.4 in] {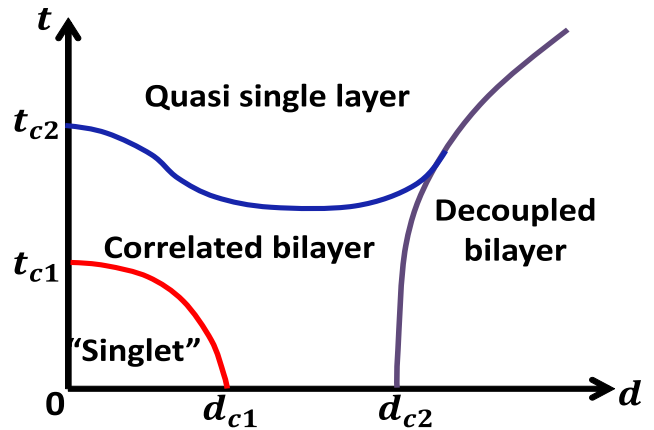}}
\caption{(Color online) Proposed "$t-d$" phase diagram for the $\nu=4/5$ bilayer system.} \label{phd2}
\end{figure}

In the end, we would like to summarize our findings in a schematic $t-d$ phase diagram, Fig. 7, for the $\nu=4/5$ bilayer system as described by the model Hamiltonian Eq.\eqref{ham-t}: The system remains  a pseudospin-singlet in the small-$t$/small-$d$ region, becomes a decoupled two $2/5$-layer system at large $d$, and a quasi-single-layer at large $t$, with a correlated bilayer of 5-fold GS degeneracy in the intermediate region.

\section{Conclusion}

We  have presented detailed exact diagonalization studies on various toroidal geometries for bilayer FQH states at filling fraction $4/5$, with up to $N_e=12$ particles. We found that for small enough layer separation $d$, the regime of two weakly coupled $2/5$-layers undergoes a phase transition into another Abelian regime, characterized by five-fold (minimal) ground state degeneracy. The state at zero layer separation has been shown to be a pseudo-spin singlet. We have discussed possible $K$-matrix descriptions for the Abelian regime with small- to intermediate-$d$, ruling out all $2\times 2$ $K$-matrices by analyzing the Chern-number matrix in this regime. In the absence of additional conservation laws, we find only one equivalence class of $4\times 4$ matrices with the proper quantum numbers. A model wave function representing this class has been constructed, consisting of two $2/3$-states coupled by an interlayer Jastrow-factor. In the presence of separate particle number conservation in each layer, our study suggests the existence of two separate symmetry protected Abelian phases with identical topological quantum numbers, in particular, identical ground state degeneracy, but different Chern-number matrix. We have further studied the fate of these phases in the presence of finite interlayer-tunneling, finding in particular that all phase boundaries survive finite tunneling, but in part become first order.
We are hopeful that these findings will stimulate further investigation of this rich and interesting regime in both theory and experiment.

\section{Acknowledgments}

H.W.'s work is supported by the National Natural Science Foundation of China (NSFC) Grant No. 11474144. K.Y.'s work is supported by DOE Grant No. DESC0002140 and performed at the National High Magnetic Field Laboratory, which is supported by National Science Foundation Cooperative Agreement No. DMR-1644779, and the State of Florida. F.C.Z's work is partially supported by the strategic priority research program of CAS Grant No. XDB28000000, and by NSFC Grant No. 11674278, and National Basic Research Program of China (No. 2014CB921203). A.S. would like to thank Meng Cheng for insightful discussions.

\end{document}